\documentclass[aps,prb,twocolumn,superscriptaddress]{revtex4-1}
\usepackage{dcolumn,amsfonts,amsmath,amssymb}
\usepackage{bm}% bold math

\usepackage{color}
\usepackage[pdftex]{graphicx}
\usepackage{epstopdf}

\newcommand{\kp}{|+\rangle}
\newcommand{\km}{|-\rangle}
\newcommand{\bp}{\langle+|}
\newcommand{\bmm}{\langle-|}
\newcommand{\ktp}{|T+\rangle}
\newcommand{\ktm}{|T-\rangle}
\newcommand{\btp}{\langle T+|}
\newcommand{\btm}{\langle T-|}
\newcommand{\kup}{|\!\uparrow\rangle}
\newcommand{\kdn}{|\!\downarrow\rangle}
\newcommand{\ktup}{|T\!\uparrow\rangle}
\newcommand{\ktdn}{|T\!\downarrow\rangle}
\newcommand{\bup}{\langle\uparrow\!|}
\newcommand{\bdn}{\langle\downarrow\!|}
\newcommand{\btup}{\langle T\!\uparrow\!|}

\newcommand{\wt}{\omega_{\mathrm{t}}}
\newcommand{\wh}{\omega_{\mathrm{h}}}
\newcommand{\St}{\Sigma_{\mathrm{t}}}
\newcommand{\Sh}{\Sigma_{\mathrm{h}}}
\newcommand{\Nt}{N_{\mathrm{t}}}

\newcommand{\Yh}{Y_{\mathrm{h}}}

\newcommand{\tr}[1]{\mathrm{Tr}\left( #1 \right)}

\begin{document}
\title{Spin dynamics in p-doped semiconductor nanostructures
subject to a magnetic field tilted from the Voigt geometry}
\author{K. Korzekwa}\email{k.korzekwa12@imperial.ac.uk}
\affiliation{Department of Physics, Imperial College London, London SW7 2AZ, United Kingdom}
\affiliation{Institute of Physics, Wroc{\l}aw University of Technology, 50-370 Wroc{\l}aw, Poland}
\author{C.\ Gradl}
\affiliation{Institut f\"ur Experimentelle und Angewandte Physik,
Universit\"at Regensburg, D-93040 Regensburg, Germany}
\author{M.\ Kugler}
\affiliation{Institut f\"ur Experimentelle und Angewandte Physik,
Universit\"at Regensburg, D-93040 Regensburg, Germany}
\author{S.\ Furthmeier}
\affiliation{Institut f\"ur Experimentelle und Angewandte Physik,
Universit\"at Regensburg, D-93040 Regensburg, Germany}
\author{M.\ Griesbeck}
\affiliation{Institut f\"ur Experimentelle und Angewandte Physik,
Universit\"at Regensburg, D-93040 Regensburg, Germany}
\author{M.\ Hirmer}
\affiliation{Institut f\"ur Experimentelle und Angewandte Physik,
Universit\"at Regensburg, D-93040 Regensburg, Germany}
\author{D.\ Schuh}
\affiliation{Institut f\"ur Experimentelle und Angewandte Physik,
Universit\"at Regensburg, D-93040 Regensburg, Germany}
\author{W.\ Wegscheider}
\affiliation{Solid State Physics Laboratory, ETH Zurich, 8093 Zurich, Switzerland}
\author{T.\ Kuhn}
\affiliation{Institut f\"ur Festk\"orpertheorie, Westf\"alische Wilhelms-Universit\"at, D-48149 M\"unster, Germany}
\author{C.\ Sch\"uller}
\affiliation{Institut f\"ur Experimentelle und Angewandte Physik,
Universit\"at Regensburg, D-93040 Regensburg, Germany}
\author{T.\ Korn}
\affiliation{Institut f\"ur Experimentelle und Angewandte Physik, Universit\"at
Regensburg, D-93040 Regensburg, Germany}
\author{P. Machnikowski}
\affiliation{Institute of Physics, Wroc{\l}aw University of
  Technology, 50-370 Wroc{\l}aw, Poland}
\date{\today}

\begin{abstract}
We develop a theoretical description of the spin dynamics of resident holes in a p-doped semiconductor quantum well (QW) subject to a magnetic field tilted from the Voigt geometry. We find the expressions for the signals measured in time-resolved Faraday rotation (TRFR) and resonant spin amplification (RSA) experiments and study their behavior for a range of system parameters. We find that an inversion of the RSA peaks can occur for long hole spin dephasing times and tilted magnetic fields. We verify the validity of our theoretical findings by performing a series of TRFR and RSA experiments on a p-modulation doped GaAs/Al$_{0.3}$Ga$_{0.7}$As single QW and showing that our model can reproduce experimentally observed signals.
\end{abstract}

\maketitle

\section{Introduction}

In recent years, it has become evident that spin dynamics of carriers in low-dimensional semiconductor structures can be studied in a controlled way by optical means~\cite{Dyakonov_Book,Korn10}. Apart from the general interest in understanding the often non-trivial kinetics of spin precession and decoherence, the optical studies of semiconductor spin dynamics are motivated by possible applications in spintronics~\cite{Fabian04,fabian07,WuReview} and spin-based quantum information processing~\cite{awschalom02}. From the point of view of these applications, the extended coherence time of holes confined in quantum dots~\cite{heiss07,Warburton09,syperek12} and localized in quantum wells (QW)~\cite{marie95,marie99,syperek07,Korn10njp}, resulting from suppression of the major spin dephasing channels in confined systems, seems to be very promising and encourages investigation of these systems, both experimental and theoretical.

Among the optical methods used in these investigations, the time-resolved magnetooptical Faraday/Kerr
rotation (TRFR/TRKR)~\cite{baumberg94} and the related resonant spin amplification (RSA)~\cite{kikkawa98} techniques have been shown to
yield particularly rich information on the coherent dynamics and dephasing of electron and hole spins. The use of spectrally narrow lasers with ps pulse length allows for selective, resonant excitation of, e.g., exciton or trion transitions~\cite{Chen07, yugova:167402}, as well as nonresonant excitation with well-defined excess energy~\cite{Studer11,Kugler11}. In TRFR/TRKR experiments, a pump-probe measurement scheme is employed, and the observable time window is typically limited to a few ns, making accurate measurements of long spin dephasing times difficult. The RSA technique utilizes  the interference of spin polarizations created in a sample by subsequent pump pulses to circumvent this limitation of TRFR. A variable in-plane magnetic field is used to induce spin precession, and the Faraday rotation angle is measured for a fixed time delay. For certain magnetic field values, constructive interference of spin polarizations occurs, giving rise to characteristic maxima in the Faraday rotation signal. Even though this RSA signal is typically recorded for a large delay between pump and probe pulses, its  shape is strongly influenced by the combined carrier and spin dynamics during the photocarrier lifetime~\cite{yugova:167402,Korn10njp,Kugler11}, and may also reflect anisotropic spin dephasing~\cite{Glazov_RSA,Griesbeck12}.

While many of the experimental results can be accounted for by relatively simple models~\cite{syperek07,Korn10njp}, full understanding of the role of various microscopic factors underlying the observed dynamics, including the g-factor anisotropy and channels of dephasing, requires a more formal theoretical modeling. Such models were proposed for n-doped \cite{yugova09} and p-doped~\cite{Kuhn10} systems, rigorously relating the observed optical response to the precession and decay of the optically induced spin polarization in the sample. It was pointed out~\cite{Kuhn10} that the optically manifested spin dynamics becomes particularly rich if the magnetic field is tilted from the in-plane (Voigt) orientation, leading to the coexistence of oscillating and exponential components in the Faraday response. The theory was subsequently extended to model the RSA response of confined holes and electrons~\cite{Kugler11,zhukov12,yugova12}. In the latter case, it was again pointed out that the shape of RSA signals is very sensitive to the magnetic field orientation deviating from the exact Voigt geometry.

Here, we present time-resolved studies of the combined electron and hole spin dynamics in a p-modulation doped QW subject to a magnetic field tilted from the Voigt geometry. Using the Markovian master equation we develop a theoretical description of the spin dynamics and find analytical expressions for signals of two widely employed magnetooptical experiments: the time-resolved Faraday rotation and the resonant spin amplification. We identify the physical processes responsible for the formation of the signals and discuss their behavior dependent on the parameters of the system. We also point out the possibility for the emergence of  inverted RSA peaks induced by a tilted magnetic field, when the spin dephasing time is long enough. Finally, we verify our theoretical findings by performing a series of experimental measurements on GaAs-based p-doped single QWs and demonstrate that the model can account for all experimentally observed features of both signals. Results presented here generalize our former findings on magnetooptical experiments in tilted magnetic fields \cite{Kuhn10} by finding the angular dependence of the hole spin decoherence rates in the TRFR experiment and extend the decription to include the RSA experiment.

The paper is organized as follows. In Sec.~\ref{sec:theor}, we introduce the theoretical model describing the TRFR and RSA responses of the p-doped QW. Next, in Sec.~\ref{sec:sample}, we present the sample used in our experiment and describe the experimental setup and methods used. Sec.~\ref{sec:results} contains the physical interpretation and discussion of the theoretical results together with the analysis of the experimental measurements based on our model. Finally, Sec.~\ref{sec:concl} concludes the paper.

\section{Theoretical model}
\label{sec:theor}

It is known that the experimentally measured Faraday signal gives access to the spin polarization of the sample at the arrival of the probe pulse \cite{Kuhn10,yugova09}. Therefore, in order to interpret the experimental observations, we focus on the underlying microscopic electron and hole spin dynamics. Basing on the previous experimental findings \cite{syperek07,kugler:035325} and our recent theoretical works \cite{Kuhn10,Kugler11} we model the system in the following way.

The optical response is assumed to come from independent hole-trion systems, trapped in QW fluctuations and restricted to the 2 lowest spin states (due to the large heavy-light hole splitting in confined systems the description is restricted to the heavy-hole states, treated as a pseudo-spin-1/2 system). Each such system is represented by a density matrix $\rho$ restricted to the four states $\kup,\kdn,\ktup,\ktdn$, representing the two hole states and the two trion states with different spin orientations (with respect to the normal to the sample plane). The density matrix is parametrized by introducing the set of dynamical variables
\begin{subequations}
\begin{eqnarray}
N_{\mathrm{h/t}}&=&\tr{\sigma_{0}^{\mathrm{(h/t)}}\rho},\label{eq:dyn_var1}\\
X_{\mathrm{h/t}}&=&\tr{\sigma_{1}^{\mathrm{(h/t)}}\rho},\\
Y_{\mathrm{h/t}}&=&\tr{\sigma_{2}^{\mathrm{(h/t)}}\rho},\\
\Sigma_{\mathrm{h/t}}&=&\tr{\sigma_{3}^{\mathrm{(h/t)}}\rho},\label{eq:dyn_var4}
\end{eqnarray}
\end{subequations}
where $\sigma_{i}^{\mathrm{(h/t)}}$ are Pauli operators restricted to hole/trion subspace ($\sigma_{0}$ is an identity operator). The Hamiltonian of the system placed in a magnetic field oriented at an angle $\theta$ with respect to the growth axis, $\mathbf{B}=B(\sin\theta,0,\cos\theta)$, and in a reference frame rotating with zero-field hole-trion transition frequency (which we will use throughout this work) is given by
\begin{eqnarray}
\label{eq:H0_1}
H_{0}=-\frac{1}{2}\mu_{\mathrm{B}}\bm{B}\hat{g}_{\mathrm{h}}\bm{\sigma}^{\mathrm{(h)}}
-\frac{1}{2}g_{\mathrm{t}}\mu_{\mathrm{B}}\bm{B}\cdot\bm{\sigma}^{\mathrm{(t)}},
\end{eqnarray}
where $\mu_{\mathrm{B}}$ is the Bohr magneton, $\hat{g}_{\mathrm{h}}$ is the hole Land{\'e} tensor and $g_{\mathrm{t}}$ is the Land{\'e} factor of the trion (i.e., of the electron), which is assumed to be isotropic. The hole Land{\'e} tensor is assumed to have no in-plane anisotropy, so it is characterized by the in-plane component $g_{\perp}$ (perpendicular to the normal to the QW plane) and the out-of-plane component $g_{\parallel}$ (parallel to the normal to the QW plane). Then, introducing the effective hole Land{\'e} factor $\tilde{g}=(g_{\perp}^2\sin^2\theta+g_{\parallel}^2\cos^2\theta)^{1/2}$ and the angle $\phi$ such that $\tan\phi=(g_{\perp}/g_{\parallel})\tan\theta$ (which is the angle between the growth axis and the hole spin quantization axis, as the latter one does not necessarily coincide with the field orientation) this Hamiltonian may be rewritten in its eigenbasis:
\begin{eqnarray}
\label{eq:H0_2}
H_{0}&=&\frac{1}{2}\hbar\wh(\km\bmm-\kp\bp)\nonumber\\
&&+\frac{1}{2}\hbar\wt(\ktm\btm-\ktp\btp),
\end{eqnarray}
where $\hbar\wh=\tilde{g}\mu_{\mathrm{B}}B$, $\hbar\wt=g_{\mathrm{t}}\mu_{\mathrm{B}}B$ and
\begin{subequations}
\begin{eqnarray*}
\kp&=&\cos\frac{\phi}{2}\kup+\sin\frac{\phi}{2}\kdn,\\
\km&=&-\sin\frac{\phi}{2}\kup+\cos\frac{\phi}{2}\kdn,\\
\ktp&=&\cos\frac{\theta}{2}\ktup+\sin\frac{\theta}{2}\ktdn,\\
\ktm&=&-\sin\frac{\theta}{2}\ktup+\cos\frac{\theta}{2}\ktdn.
\end{eqnarray*}
\end{subequations}

Having described the system we now proceed to the description of its evolution. This is divided into two steps: first, the driven evolution under the pump pulse, and then the free evolution treated in an open quantum system formalism, i.e. Larmor precession, recombination and spin decoherence.

The coupling of the hole and trion states by the circularly polarized ($\sigma^+$) laser field is treated in the dipole approximation and the corresponding Hamiltonian in the rotating wave approximation is given by
\begin{eqnarray}
H_{\mathrm{l}}&=&\frac{1}{2}f(t)\mbox{$\kup\!\btup$}+\mathrm{h.c.},
\end{eqnarray}
where $f(t)$ is the pulse envelope function. Assuming that the pump pulse is low-power and short enough, i.e. much shorter than any relevant time scale of the system dynamics, the transformation it induces is described as instantaneous and up to the second order in the pulse amplitude,
\begin{eqnarray}
\rho_{1} & = &
-\frac{i}{\hbar}\int_{-\infty}^{\infty} dt
\left[H_{\mathrm{l}}(t),\rho_{0}\right] \nonumber \\
&&-\frac{1}{\hbar^{2}}\int_{-\infty}^{\infty}dt\int_{-\infty}^{\infty}dt'
\left[H_{\mathrm{l}}(t),
  \left[H_{\mathrm{l}}(t'),\rho_{0}\right]\right],
\label{eq:laser}
\end{eqnarray}
where $\rho_0$ is the initial state of the system and $\rho_1$ is the state just after the arrival of the pump pulse. For the TRFR experiment $\rho_0$ is the thermal equilibrium state $\rho_{\mathrm{eq}}$ (since the evolution after each laser repetition is independent), for which all trion variables are zero and the hole variables are parametrized by equilibrium spin polarization along hole spin quantization axis,
\begin{equation*}
p=\bp\rho_{\mathrm{eq}}\kp-\bmm\rho_{\mathrm{eq}}\km=
\tanh\left( \frac{\hbar\wh }{2k_{\mathrm{B}}T} \right),
\end{equation*}
so that $\Sigma_{\mathrm{h}}^{\mathrm{(eq)}}=p\cos\phi$ and $X_{\mathrm{h}}^{\mathrm{(eq)}}=p\sin\phi$.

The free evolution in between the pump and probe pulses is modeled using the Markovian master equation,
\begin{equation}\label{eq:evol}
\dot{\rho}=-\frac{i}{\hbar}[H_{0},\rho]+\mathcal{L}[\rho],
\end{equation}
with the initial condition given by $\rho_1$. In Eq. (\ref{eq:evol}) the first term accounts for the spin precession and the dissipative dynamics is described by universal Lindblad superoperator $\mathcal{L}=\mathcal{L}_{\mathrm{h}}+\mathcal{L}_{\mathrm{r}}$. Here $\mathcal{L}_{\mathrm{h}}$ is the hole spin dissipator, $\mathcal{L}_{\mathrm{r}}$ is the spontaneous emission generator and no trion spin dissipator is included as it is assumed that the radiative decay rate is much larger than any trion decoherence rates.

The hole spin dissipator, $\mathcal{L}_{\mathrm{h}}$, is obtained using the standard weak-coupling approach \cite{breuer02} from the hole spin-environment Hamiltonian
\begin{equation}
\label{eq:H_int}
H_{\mathrm{int}}=\sigma_+^{(\mathrm{h})}R_-^{(\mathrm{h})}+\sigma_-^{(\mathrm{h})}R_+^{(\mathrm{h})}+\sigma_{0}^{(\mathrm{h})}R_0^{(\mathrm{h})},
\end{equation}
where $\sigma_+^{(\mathrm{h})}=\left(\sigma_-^{(\mathrm{h})}\right)^{\dagger}=\kp\bmm$ and the environment operators are definied by
\begin{subequations}
\begin{eqnarray*}
\label{eq:envi_op}
R_-^{(\mathrm{h})}&=&R_x^{(\mathrm{h})}\cos\phi-iR_y^{(\mathrm{h})}-R_z^{(\mathrm{h})}\sin\phi,\\
R_+^{(\mathrm{h})}&=&R_x^{(\mathrm{h})}\cos\phi+iR_y^{(\mathrm{h})}-R_z^{(\mathrm{h})}\sin\phi,\\
R_0^{(\mathrm{h})}&=&R_x^{(\mathrm{h})}\sin\phi+R_z^{(\mathrm{h})}\cos\phi.
\end{eqnarray*}
\end{subequations}
The above environmental operators are expressed in terms of operators defined by the system structure to allow us to use the system symmetry in order to achieve considerable simplifications (see below). Using this spin-environment Hamiltonian the following dissipator is obtained
\begin{eqnarray}
\label{eq:L_h}
\mathcal{L}_{\mathrm{h}}[\rho]&=& -\pi\sum_{lj}\left[\vphantom{\left(R_{lj}^{(\mathrm{h})}\right)}
R_{lj}^{(\mathrm{h})} (\omega_{j})\left( \sigma_{l}^{(\mathrm{h})}\sigma_{j}^{(\mathrm{h})}\rho
-\sigma_{j}^{(\mathrm{h})}\rho\sigma_{l}^{(\mathrm{h})} \right)\right.\nonumber\\ &&\left.+R_{lj}^{(\mathrm{h})} (-\omega_{l})\left( \rho  \sigma_{l}^{(\mathrm{h})}\sigma_{j}^{(\mathrm{h})}-\sigma_{j}^{(\mathrm{h})}\rho\sigma_{l}^{(\mathrm{h})} \right)\right]  ,
\end{eqnarray}
where $l,j=\pm,0$, $\omega_{0}=0$, $\omega_{+}=-\omega_{-}=\omega_{\mathrm{h}}$ and the spectral densities for the hole reservoir are defined as:
\begin{equation*}
R_{lj}^{(\mathrm{h})} (\omega)=\frac{1}{2\pi\hbar^{2}}\int dt e^{i\omega t}
\langle R_{l}^{(\mathrm{h})} (t)R_{j}^{(\mathrm{h})}\rangle,
\quad l,j=\pm,0,
\end{equation*}
Now the forementioned simplification of spectral densities can be obtained by assuming the system $C_{4\mathrm{v}}$ symmetry and setting $R_{\alpha\beta}^{(\mathrm{h})}(\omega)=0$ for $\alpha,\beta=x,y,z$,$\alpha\neq\beta$ and $R_{yy}^{(\mathrm{h})}(\omega)=R_{xx}^{(\mathrm{h})}(\omega)$.

The spontaneous emission generator, $\mathcal{L}_{\mathrm{r}}$, accounting for the radiative recombination of the trion, has a standard form,
\begin{eqnarray}
\label{eq:L_r}
L_{\mathrm{r}}[\rho] & = & \gamma\left[
\sigma_{-}^{(\uparrow)}\rho\sigma_{+}^{(\uparrow)}
-\frac{1}{2}\left\{
\sigma_{+}^{(\uparrow)}\sigma_{-}^{(\uparrow)},\rho\right\}_{+}\right.\nonumber\\
&&+\left.\sigma_{-}^{(\downarrow)}\rho\sigma_{+}^{(\downarrow)}
-\frac{1}{2}\left\{
\sigma_{+}^{(\downarrow)}\sigma_{+}^{(\downarrow)},\rho\right\}_{+}
\right]
\end{eqnarray}
where $\gamma$ is the radiative decay rate, \mbox{$\sigma_{+}^{(\uparrow)}=\left( \sigma_{-}^{(\uparrow)}\right)^{\dag} = \ktup\bup$} and \mbox{$\sigma_{+}^{(\downarrow)}=\left( \sigma_{-}^{(\downarrow)}\right)^{\dag}=\ktdn\bdn$}. Note that no pure dephasing rate is included, as trion coherence contributes neither to the TRFR nor to the RSA signal.

This derivation is based on the one presented in Ref. \onlinecite{Kuhn10}, however with a significant difference. Namely, no secular approximation is used for the spin dynamics, so the description for arbitrary Larmor frequencies (magnetic fields) is obtained. Although later we use some approximations based on the the magnitude of the magnetic field, omitting this kind of approximation at the level of the equation of motion derivation preserves the angular dependence of decoherence rates, which is otherwise lost.

The equation of motion for the density matrix, Eq. (\ref{eq:evol}), can be rewritten in terms of the dynamical variables defined by Eqs. (\ref{eq:dyn_var1})-(\ref{eq:dyn_var4}). Differential equations for trion variables obtained in this way can easily be solved and the solutions for trion population, $\Nt$, and spin polarization, $\St$, are given by
\begin{subequations}
\begin{eqnarray}
\Nt(t)&=&\Nt(0)e^{-\gamma t},\\
\St(t)&=&\St(0)e^{-\gamma t}\label{eq:sol_t}
\left( \cos^{2}\theta+\sin^{2}\theta\cos\wt t \right).
\end{eqnarray}
\end{subequations}
The dynamics of the hole variables is governed by the following set of differential equations
\small
\begin{subequations}
\begin{eqnarray}
\dot{\widetilde{\Sigma}}_{\mathrm{h}} & =& -\left(\frac{3+\cos 2\phi}{2}\kappa_{x} +\frac{1-\cos 2\phi}{2}\kappa_{x0}
\right)\widetilde{\Sigma}_{\mathrm{h}}\nonumber\\
&&-\left(\frac{\kappa_{x}-\kappa_{x0}}{2} \right)\sin 2 \phi \widetilde{X}_{\mathrm{h}}-\wh \sin\phi \Yh\nonumber\\
&&-2\kappa '_{x}\cos\phi \Nt+\gamma \St,\label{eq:diff_sh}\\\nonumber&&\\
\dot{\widetilde{X}}_{\mathrm{h}} & =&
-\left( \frac{1-\cos 2\phi}{2}\kappa_{z}
+\frac{1+\cos 2\phi}{2}\kappa_{z0} + \kappa_{x}
 \right) \widetilde{X}_{\mathrm{h}}\nonumber\\
&&- \frac{\kappa_{z}-\kappa_{z0}}{2}\sin 2\phi \widetilde{\Sigma}_{\mathrm{h}} +\wh\cos\phi \Yh\nonumber\\
&&+(\kappa '_{x}-\kappa '_{z})\sin\phi \Nt,\label{eq:diff_xh}\\\nonumber&&\\
\dot{Y}_{\mathrm{h}} & = & -\left(
\frac{1+\cos 2\phi}{2}\left( \kappa_{z0}+\kappa_{x} \right)
+\frac{1-\cos 2\phi}{2}\left( \kappa_{x0} +\kappa_{z} \right)
 \right) \Yh\nonumber\\
&&+\wh \sin\phi\widetilde{\Sigma}_{\mathrm{h}} -\wh\cos\phi \widetilde{X}_{\mathrm{h}},\label{eq:diff_yh}
\end{eqnarray}
\end{subequations}\normalsize
where new hole variables $\widetilde{\Sigma}_{\mathrm{h}}$, $\widetilde{X}_{\mathrm{h}}$ with subtracted equilibrium values ($\Sigma_{\mathrm{h}}^{\mathrm{(eq)}}$ and $X_{\mathrm{h}}^{\mathrm{(eq)}}$) are used and the decoherence rates (for $\alpha=x,z$) are
\begin{subequations}
\begin{eqnarray}
\label{eq:kappa1}
\kappa_{\alpha} & = &2\pi \left[R_{\alpha\alpha}(\omega_{\mathrm{h}})+R_{\alpha\alpha}(-\omega_{\mathrm{h}}) \right], \\
\kappa'_{\alpha} & =& 2\pi \left[R_{\alpha\alpha}(\omega_{\mathrm{h}})-R_{\alpha\alpha}(-\omega_{\mathrm{h}}) \right], \\
\kappa_{\alpha} & = & 4\pi R_{\alpha\alpha}(0).\label{eq:kappa3}
\end{eqnarray}
\end{subequations}
In order to find the physical meaning of these decoherence rates we consider 3 limiting situations. For $\mathbf{B}=0$ the decoherence time for the out-of-plane component of spin polarization is $T_z^{(0)}=1/2\kappa_{x0}$ and for the in-plane component it is $T_{xy}^{(0)}=1/(\kappa_{x0}+\kappa_{z0})$. In the case of strong in-plane magnetic field (\mbox{$\wh\gg\kappa_{\alpha},\kappa '_{\alpha},\kappa_{\alpha 0}$} and $\theta=\phi=\pi/2$) the spin relaxation time $T_1$ and the spin dephasing time $T_2$ are given by
\begin{subequations}
\begin{eqnarray}
\label{eq:t1t2x_1}
T_1^{(x)}&=&\frac{1}{\kappa_x+\kappa_z}\equiv\frac{1}{\kappa_{1\perp}},\\
T_2^{(x)}&=&\frac{2}{\kappa_x+\kappa_z+2\kappa_{x0}}\equiv\frac{1}{\kappa_{2\perp}}.\label{eq:t1t2x_2}
\end{eqnarray}
\end{subequations}
For the strong out-of-plane magnetic field (\mbox{$\wh\gg\kappa_{\alpha},\kappa '_{\alpha},\kappa_{\alpha 0}$} and $\theta=\phi=0$) these times are given by:
\begin{subequations}
\begin{eqnarray}
\label{eq:t1t2z_1}
T_1^{(z)}&=&\frac{1}{2\kappa_x}\equiv\frac{1}{\kappa_{1\parallel}},\\
T_2^{(z)}&=&\frac{1}{\kappa_x+\kappa_{z0}}\equiv\frac{1}{\kappa_{2\parallel}}.\label{eq:t1t2z_2}
\end{eqnarray}
\end{subequations}

Basing on the results from Refs. \onlinecite{Kuhn10,yugova09}, we connect the rotation of the polarization plane of the transmitted probe pulse with the difference of hole and trion spin polarization,
\begin{equation}
\Delta\Sigma=\St-\Sh.
\end{equation}

In order to model the TRFR experiment we analytically solve the set of differential equations, Eqs. (\ref{eq:diff_sh})-(\ref{eq:diff_yh}), and approximate the final solution for $\Sh$ by assuming that hole decoherence rates are smaller than the other dynamical parameters of the system. This approximation is plausible, as TRFR experiments are usually performed in magnetic fields for which the precession period is much shorter than the hole spin decoherence time. In this way we obtain the following expression for the Faraday signal,
\begin{eqnarray}
\Delta\Sigma^{\mathrm{(Faraday)}}&=&\St(t)-\left(A_1 e^{-\gamma t}+A_2 e^{-(\gamma+i\wt)t}\right.~~~~\nonumber\\
&&\left.+B_1e^{-\kappa_1t}+B_2 e^{-(\kappa_2+i\wh)t}+\mathrm{c.c.}\right)\!\!,\label{eq:TRKR_tilt}
\end{eqnarray}
where $\St(t)$ is given by Eq. (\ref{eq:sol_t}), new decoherence rates are
\begin{subequations}
\begin{eqnarray}
\label{eq:kappa_eff1}
\kappa_1&=&\kappa_{1\perp}\sin^2\phi+\kappa_{1\parallel}\cos^2\phi,\\
\kappa_2&=&\kappa_{2\perp}\sin^2\phi+\kappa_{2\parallel}\cos^2\phi,\label{eq:kappa_eff2}
\end{eqnarray}
\end{subequations}
and
\small
\begin{eqnarray*}
A_1&=&-\frac{1}{2}\frac{\wh^{2}\cos^{2}\phi+\gamma^{2}}{\wh^{2}+\gamma^{2}}\cos^{2}\theta\St(0),\label{eq:A1_tilt}\\
A_2&=&-\frac{1}{2}\frac{\gamma}{\gamma+i\wt}\frac{\wh^{2}\cos^{2}\phi+(\gamma+i\wt)^{2}}{\wh^{2}+(\gamma+i\wt )^{2}}\sin^{2}\theta \St(0),\label{eq:A2_tilt}\\
B_1&=&\frac{1}{2}\frac{\wt^{2}\cos^2\theta+\gamma^2}{\wt^{2}+\gamma^{2}}\cos^{2}\phi\St(0)
+\frac{1}{2}\cos^2\phi\widetilde{\Sigma}_{\mathrm{h}}(0)\nonumber\\
&&+\frac{1}{4}\sin 2\phi \widetilde{X}_{\mathrm{h}}(0),\label{eq:B1_tilt}\\
B_2&=&\frac{1}{2}\frac{\gamma}{\gamma-i\wh}\frac{\wt^{2}\cos^{2}\theta+(\gamma-i\wh)^{2}}{
\wt^{2}+(\gamma-i\wh)^{2}}\sin^{2}\phi\St(0)\nonumber\\
&&+\frac{1}{2}\sin^{2}\phi\widetilde{\Sigma}_{\mathrm{h}}(0)-\frac{1}{4}\sin 2\phi \widetilde{X}_{\mathrm{h}}(0)+\frac{i}{2}\sin\phi \Yh(0)\label{eq:B2_tilt}.
\end{eqnarray*}
\normalsize

In the RSA experiment the system repetitively undergoes the two-step evolution described before, with the time period $t_{\mathrm{r}}$ given by the repetition rate of the pump laser. The experiment is usually performed in the long spin dephasing time (SDT) regime, i.e. SDT is longer than $t_{\mathrm{r}}$, so the spin polarization surviving between subsequent laser repetitions is essential. Therefore, in order to model the RSA response, the fixed point of the two-step transformation (pump pulse and Lindblad evolution during the repetition interval $t_{\mathrm{r}}$) of the dynamical variables is found. A series of approximations is also applied in order to obtain a concise expression for the stationary spin polarizations. First of all, it is assumed that only hole spin polarization contributes to the RSA signal, as  $t_{\mathrm{r}}$ is much longer than the radiative decay time and no trion population survives until the arrival of the probe pulse. Next, the assumption is made that the hole spin decoherence rates are small compared to the trion recombination rate, which is a reasonable assumption taking into account the long SDT experimental conditions. It is also assumed that the hole Larmor frequency is larger than the decoherence rates, which requires justification. To justify this approximation, let us note that the first RSA peak occurs for $\wh=1/t_{\mathrm{r}}$ and also, due to the long SDT regime, $t_{\mathrm{r}}<\tau$, where $\tau$ stands for the effective hole decoherence time. Hence, the only discrepancies between the modeled and measured signal introduced by this approximations can occur between zero magnetic field and the first RSA peak. It is also assumed that $p\ll 1$, i.e. that the experiment is done in the high-temperature regime in the sense $\hbar\wh\ll k_{\mathrm{B}}T$, which is the usual case for the range of magnetic fields used in the RSA experiments, as the temperature in experiments varies from hundreds of millikelvins to a few kelvins. Finally, we limit our considerations to magnetic fields slightly tilted from the Voigt geometry for which $\cos^2\theta\ll 1$. To understand the physical meaning of this approximation, let us recall that $\theta$ defines the quantization axis for trion spins. However, trion spins do not contribute to the RSA signal, their only contribution is to remove some of the hole polarization during recombination\cite{syperek07}. Neglecting $\cos^2\theta$ for small tilt angles is thus equivalent to assuming that trions precessing around the slightly tilted axis remove on average (during their lifetime) holes of approximately the same polarization as without tilting. On the other hand, due to strong anisotropy of the hole $g$-factor, even small tilting results in  hole spin precession around a significantly different axis, which is accounted for by preserving terms proportional to $\cos^2\phi$. The expression for the RSA signal resulting from the procedure described above is given by
\begin{align}
\label{eq:RSA_tilt}
\Delta\Sigma^{\mathrm{(RSA)}}=&\frac{\sum_{i=1}^3A_ie^{-\lambda_it_{\mathrm{r}}}-\sum_{i=4}^5A_ie^{-\lambda_it_{\mathrm{r}}}+\mathrm{c.c.}}{Q\left(\left(\sum_{i=1}^3e^{-\lambda_it_{\mathrm{r}}}-\sum_{i=4}^5e^{-\lambda_it_{\mathrm{r}}}+\mathrm{c.c.}\right)-1\right)},
\end{align}
where
\begin{align}
\lambda_1&=\kappa_1, & \lambda_2&=\kappa_1+2\kappa_2, & \lambda_3&=\kappa_2+i\wh,\nonumber\\
\lambda_4&=2\kappa_2, & \lambda_5&=\kappa_1+\kappa_2+i\wh,&\nonumber
\end{align}
and\small
\begin{eqnarray*}
A_1&=&\cos^2\phi\wt^2((\wh+\wt)^2+\gamma^2)((\wh-\wt)^2+\gamma^2)(\wh^2+\gamma^2),\\
A_2&=&(\wh^2+\gamma^2)\left[((\wh^2-\wt^2)^2+\gamma^2(\wh^2+\wt^2))(\wt^2+\gamma^2)\right.\\
&&\left.-\cos^2\phi\wh^2\gamma^2(\wh^2-3\wt^2+\gamma^2)\right],\\
A_3&=&\sin^2\phi(\wt^2+\gamma^2)(\wh^2+\gamma^2)((\wh-i\gamma)^2-\wt^2)\times\\
&&(\wh^2+i\gamma\wh-\wt^2),\\
A_4&=&\sin^2\phi(\wt^2+\gamma^2)(\wh^2+\gamma^2)\left((\wh^2-\wt^2)^2+\gamma^2(\wh^2+\wt^2)\right),\\
A_5&=&((\wh-i\gamma)^2-\wt^2)(\wh^2+\gamma^2)\left[(\wh^2+i\gamma\wh-\wt^2)\times\right.\\
&&(\wt^2+\gamma^2)+\cos^2\phi(\wt^2\wh^2-\wh^2\gamma^2+3i\wt^2\gamma\wh\\
&&\left.-i\gamma^3\wh-\wt^4-\gamma^2\wt^2)\right],\\
Q&=&(\wt^2+\gamma^2)(\wh^2+\gamma^2)((\wt-\wh)^2+\gamma^2)((\wt+\wh)^2+\gamma^2).
\end{eqnarray*}
\normalsize

\section{Sample structure and experimental methods}
\label{sec:sample}
All measurements are performed on samples containing a single-side p-modulation-doped
GaAs/Al$_{0.3}$Ga$_{0.7}$As QW with a width of 4~nm grown by molecular beam epitaxy (MBE). The two-dimensional hole system (2DHS)
in the QW has a hole density $p = 1.1 \times 10^{11}$~cm$^{-2}$ and mobility
$\mu = 1.3 \times 10^{4}$~cm$^2/$Vs (measured at 1.3~K). To allow for measurements in transmission,  the samples are first glued to a sapphire subtrate and then thinned by mechanical grinding followed by  selective wet etching. The  sample structure contains a short-period GaAs/AlGaAs superlattice,
which serves as an etch stop.

 A pulsed Ti-Sapphire laser system generating pulses with a length of
1~ps and a spectral width of about 2~meV is used for the optical measurements. The laser energy is tuned to resonantly excite the trion transition in our sample. The repetition rate of the laser system is 80~MHz,
corresponding to a time interval of $t_r=12.5$~ns between subsequent pulses. The laser pulses are split into a circularly-polarized pump
beam and a linearly-polarized probe beam by a beam splitter. A mechanical delay line is used to create a variable time delay between
pump and probe. Both beams are focused to a diameter of about 80~$\mu$m on the sample using an achromat.

In the TRFR and RSA experiments, the circularly-polarized pump beam is
generating electron-hole pairs in the QW, with spins aligned parallel
or antiparallel to the beam direction, i.e., the QW normal,
depending on the helicity of the light.  In the
TRFR measurements, the spin polarization created perpendicular to the
sample plane by the pump beam is probed by the time-delayed probe
beam via the Faraday effect: the axis of linear polarization of the probe
beam is rotated by a small angle, which is proportional to the
out-of-plane component of the spin polarization
\cite{Kuhn10,yugova09}.   This small angle is detected using an optical bridge. A lock-in scheme is used to increase
sensitivity. The RSA technique is based on the interference of spin polarizations created in a sample by subsequent
  pump pulses. It requires that the spin dephasing time is comparable to the time delay between pump pulses. For certain magnetic fields applied in the sample plane, the optically oriented spin polarization precesses by an integer multiple of $2\pi$ in the time window between subsequent pump pulses, so that constructive
interference occurs. This leads to pronounced maxima in the Faraday rotation angle  measured for a fixed time delay as a function of the applied  magnetic field.  In our measurements, the time delay is chosen to probe the spin polarization remaining within the sample 100~ps before the arrival of a pump pulse.

Both, RSA and TRFR measurements are performed in
an optical cryostat with $^3$He insert, allowing us to lower the sample
temperatures below 400~mK and to apply magnetic fields of up to 11.5~Tesla. Here, the samples are cooled by cold $^3$He gas. The samples are mounted on a sample rod within the cryostat and can be rotated manually with respect to the magnetic field orientation. The rotation angle is measured with high precision using a laser pointer mounted to the sample rod. As all measurements are performed in transmission geometry, the sample can be rotated without any changes to the optical beam path.
\section{Results and Discussion}
\label{sec:results}

In this section, we present the physical interpretation of the obtained expressions for the TRFR and RSA signals and discuss their behavior in different limits of the model parameters. We also present experimental results and show that our modeling is able to reproduce experimentally measured TRFR and RSA signals in tilted fields. We first discuss the results for TRFR and then proceed to RSA.

\subsection{Time-resolved Faraday rotation}
\label{sec:results-kerr}

The physical meaning of the terms appearing in the expression for the Faraday response in a tilted magnetic field, Eq. (\ref{eq:TRKR_tilt}), will now be explained. First of all, since in a tilted magnetic field, the quantization axis for holes (trions) forms the angle $\phi$ ($\theta$) with the structure axis, a non-zero component of the optically oriented hole (trion) spin polarization along the quantization axis exists. Therefore, the spin polarization should split into two parts: non-precessing along the quantization axis (which coincides with magnetic field axis for the electrons, but, due to the strongly anisotropic hole $g$-factor, does not for the holes) and precessing, perpendicular to this axis. This is reflected in the analyzed Eq. (\ref{eq:TRKR_tilt}): the precessing parts of both trion and hole polarizations [coefficients $A_2$, $B_2$ and Eq. (\ref{eq:sol_t})] are proportional to $\sin^2\theta$ or $\sin^2\phi$ and the non-precessing components [coefficients $A_1$, $B_1$ and Eq. (\ref{eq:sol_t})] are proportional to $\cos^2\theta$ or $\cos^2\phi$. Secondly, let us note that Eq. (\ref{eq:TRKR_tilt}) can be divided into the short-living component, decaying with the recombination rate $\gamma$, and the long-living component, with its decay rate proportional to the hole spin decoherence rates. This results from the fact that, due to the difference in precession frequencies $\wh$ and $\wt$, during recombination not only  optically oriented holes are removed, but also  resident holes. Therefore, the trion spin polarization and part of the hole spin polarization decays on the time scale of the recombination time (the short-living component), whereas the other part of the hole spin polarization survives the recombination process (the long-living component). This mechanism of creating long-living spin polarization was first described in Ref. \onlinecite{syperek07}.

Now let us focus on the dependence of the modelled Faraday response on the magnetic field tilt angle. Since the electron $g$-factor is assumed isotropic, the trion precession frequency, $\wt$, is not affected by the tilting. However, due to the anisotropy of the hole $g$-factor, the hole precession frequency, $\wh$, changes. As the out-of-plane component of the hole $g$-factor $g_{\parallel}$ in the considered structure is larger than the in-plane component $g_{\perp}$, tilting the magnetic field from the Voigt geometry results in an increase of the hole precession frequency. This does not only shorten the period of the Faraday signal oscillation, but also changes the dephasing parameters, since they depend on the spectral densities of the reservoir at the frequency $\wh$ [see Eqs. (\ref{eq:kappa1})-(\ref{eq:kappa3})]. As already mentioned, the short-living component of the Faraday signal decays with the angle-independent rate $\gamma$, however the long-living component decays with rates $\kappa_1$ (non-precessing part) and $\kappa_2$ (precessing part). By comparing the expressions for these rates, Eqs. (\ref{eq:kappa_eff1})-(\ref{eq:kappa_eff2}), with the expressions for $T_1$ and $T_2$ times in the limiting situations, Eqs.~(\ref{eq:t1t2x_1})-(\ref{eq:t1t2x_2}) and (\ref{eq:t1t2z_1})-(\ref{eq:t1t2z_2}), one can find the following. The decay rate of the non-precessing component in the tilted field (inverse of the effective $T_1$ time) is a weighted average of the decay rates for magnetic fields applied only along $x$ (inverse of $T_1^{(x)}$) and only along $z$ (inverse of $T_1^{(z)}$), with weights being the $x$ and $z$ components of the unit vector parallel to quantization axis. The same holds for the precessing component and the corresponding effective $T_2$ time.
\begin{figure}[t!]
	\begin{center}
		\includegraphics[width=\columnwidth]{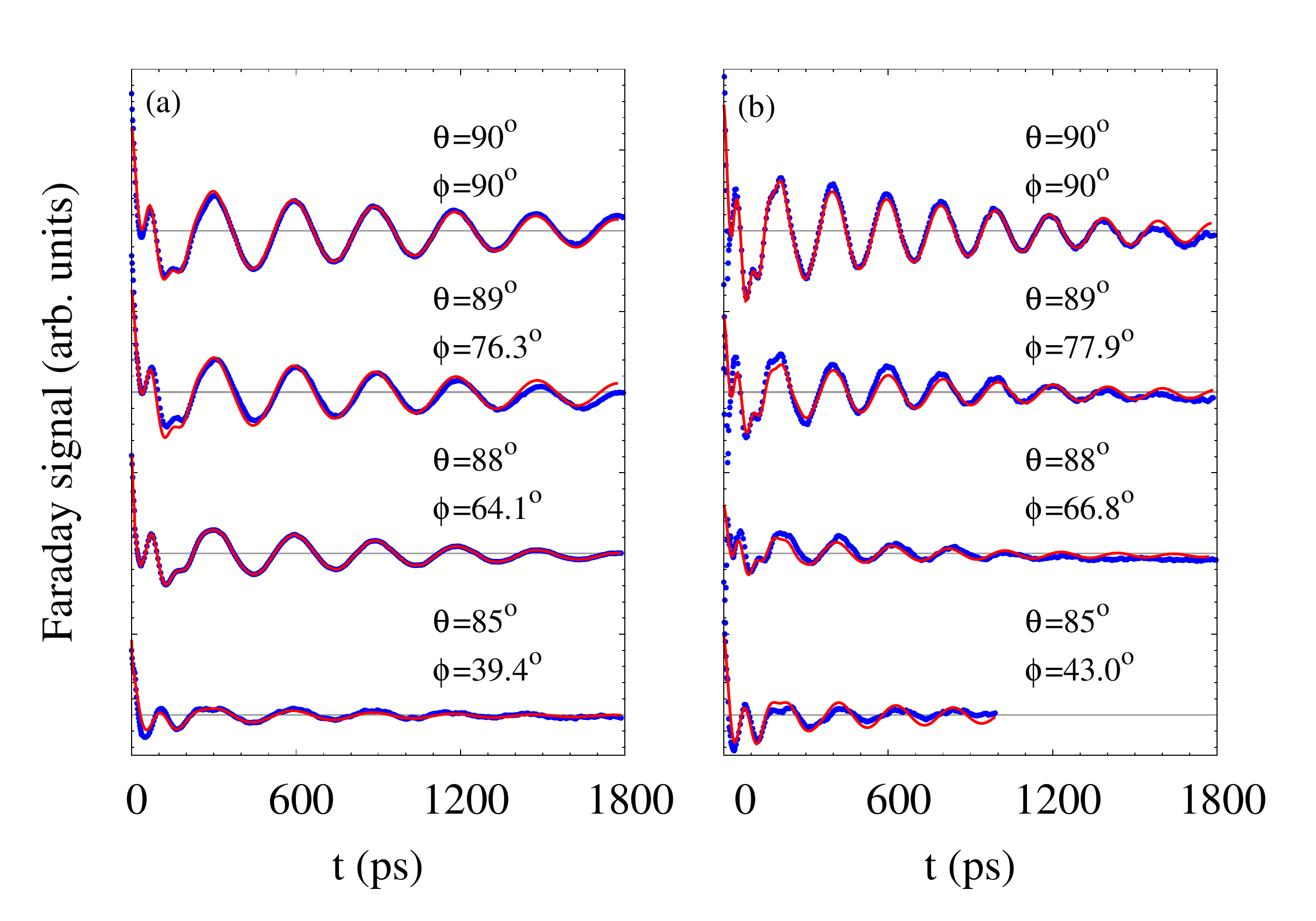}
		\caption{\label{fig:TRKR_tilt}Experimental (black dots) and modeled (red lines) Faraday signals in tilted magnetic fields for different precession periods: (a) $\tau_1\approx 300$ ps and (b) $\tau_2\approx 200$ ps.}
	\end{center}
\end{figure}

In order to verify that our model correctly describes the TRFR experiment, we performed a series of measurements. Due to the described dependence of the hole dephasing parameters on magnetic field tilt angle, the TRFR measurements were performed in the following way. Simultaneously with the increase of magnetic field tilt angle, the field amplitude  was decreased, in order to keep the hole precession frequency $\wh$ constant. In this way, the dephasing rates were kept constant, which allowed us to use the same fitting parameters ($\kappa_{1\parallel},\kappa_{1\perp},\kappa_{2\parallel},\kappa_{2\perp}$) for different tilt angles, so that the effective $T_1$ and $T_2$ times were known functions of the angle $\phi$. Two series of experimental measurements in tilted magnetic fields were performed, for two different hole precession frequencies. These corresponded to two values of the in-plane (zero tilt angle) magnetic field, $B_1=3.5$ T and $B_2=5$ T, and precession periods ($\tau=2\pi/\wh$), $\tau_1\approx 300$ ps and $\tau_2\approx 200$ ps. In each series, the tilt angle was changed from $0^{\mathrm{o}}$ ($\theta=\phi=\pi/2$) up to $5^{\mathrm{o}}$ and the temperature was set to $T=1.2$ K. The experimental data and the corresponding fitted curves, for both series of measurements, are shown in Fig. \ref{fig:TRKR_tilt}. The fitting parameters used for all angles in a given series were the same. The difference in the $\phi$ angles for the same $\theta$ angles between two series comes from the difference of the fitted out-of-plane component of the hole $g$-factor (which in turn results from the fact that in the two  experimental series, different positions on the sample with slightly different QW thickness and corresponding changes in $g_{\parallel}$ are investigated). Although the experimental traces are well-reproduced by Eq. (\ref{eq:TRKR_tilt}), there are too many free fitting parameters to make the procedure definite and infer the values of dephasing rates with certainty, i.e. the range of parameters for which the experimental data is reproduced is wide.
Also, the observed decay of the signal may come not only from the intrinsic dephasing, but also from the inhomogeneous broadening (spread of the hole $g$-factors). Hence, we conclude that our model is able to account for all the processes responsible for the formation of the Faraday signal, however in order to get quantitative insight into the investigated system spin dynamics, additional information is necessary. One needs either the information about the hole $g$-factors, their spread and the rough range of $T_1$ and $T_2$ times for different field orientation ($x$, $z$) from independent experiments or use a specific model describing hole spin decoherence and thus calculate the dephasing rates.

\subsection{Resonant spin amplification}
\label{sec:results-rsa}

First of all, let us note that the RSA signal, similarly to the TRFR response, is built up from contributions of the non-precessing spin polarization along the quantization axis and the precessing one, perpendicular to that axis. This is reflected in Eq. (\ref{eq:RSA_tilt}) by terms decaying with decay rate $\kappa_1$ (inverse of effective $T_1$ time) and $\kappa_2$ (inverse of effective $T_2$ time) being proportional to $\sin^2\phi$ and $\cos^2\phi$, respectively (exponents with $\lambda_1$, $\lambda_3$ and $\lambda_4$ factors). Due to the resonant character of the signal formation process there are also mixed components, with a decay rate being the superposition of $\kappa_1$ and $\kappa_2$ (exponents with $\lambda_2$ and $\lambda_5$ factors), that do not vanish for any angle.

\begin{figure}[t!]
	\begin{center}
		\includegraphics[width=\columnwidth]{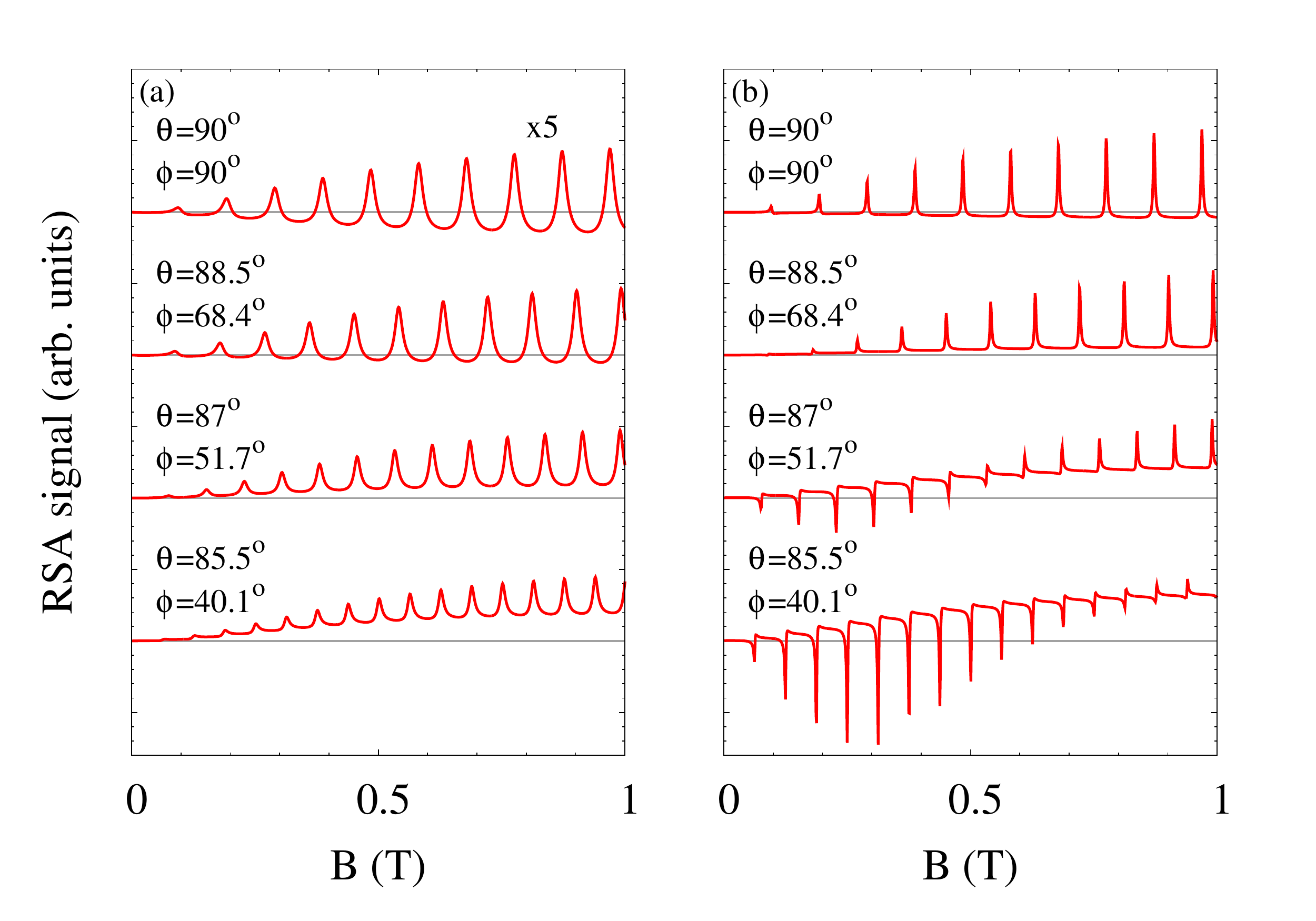}
		\caption{\label{fig:RSA_model}Modeled RSA signals for tilted magnetic fields. The parameters of the spin dynamics chosen for GaAs/AlGaAs QW \cite{Korn10}: $g_{\mathrm{t}}=0.266$, $g_{\perp}=0.059$, $g_{\parallel}=0.89$, $\gamma=10^{10}$ s$^{-1}$. (a) Standard decoherence regime: $\kappa_x=\kappa_{x0}=3\cdot 10^7$ s$^{-1}$, $\kappa_z=\kappa_{z0}=1.5\cdot 10^7$ s$^{-1}$ [signals multiplied by 5 compared to the ones presented in the (b) panel]; (b) Weak decoherence regime: $\kappa_x=\kappa_{x0}=6\cdot 10^6$ s$^{-1}$, $\kappa_z=\kappa_{z0}=3\cdot 10^6$ s$^{-1}$.}
	\end{center}
\end{figure}

Now, in order to analyze the signal dependence on the magnetic field tilt angle, we shall divide the system parameter space into two regions. In the first case, which we will refer to as standard decoherence regime, we assume that the hole spin decoherence rates are strong enough that the main contribution to the denominator of Eq. (\ref{eq:RSA_tilt}) comes from the constant $(-1)$ term. Then, the oscillations coming from the denominator are negligible and the signal shape (its dependence on magnetic field, i.e. $\wh$) is ruled by the numerator. In the second case, the weak decoherence regime, we assume that the hole spin decoherence rates are so small, that approximately $\exp(-\lambda_1t_{\mathrm{r}})\approx\exp(-\lambda_2t_{\mathrm{r}})\approx\exp(-\lambda_4t_{\mathrm{r}})\approx 1$. Then, the denominator is purely oscillatory and the overall signal shape comes from complex interplay of numerator and denominator oscillations. The exemplary dependences of the RSA signal on the magnetic field tilt angle for the standard and weak decoherence regimes are shown in Fig. \ref{fig:RSA_model}(a) and \ref{fig:RSA_model}(b), respectively.

In both regimes, one can observe a decrease in the spacings of the RSA peaks. This simply results from the increase of the hole Larmor frequency, as the effective g-factor grows with the field tilting (due to the out-of-plane component of the g-factor being much larger than the in-plane component). Another behavior, which is common for both regimes, is the tilting-induced positive drift of the average value of the signal with growing magnetic field. This can be understood in the following way. As mentioned earlier, the optically oriented spin polarization created by the pump pulse can be decomposed into two components: non-precessing along the quantization axis and precessing, perpendicular to it. As part of the non-precessing component survives in between the subsequent pump pulses, the RSA signal oscillates around its value with amplitude proportional to the surviving precessing part. Thus, together with the positive drift, the amplitude of oscillations becomes smaller. The drift grows with growing magnetic field, because trions with a larger Larmor frequency during recombination remove holes with spin directions that are more uniformly distributed, not only the optically oriented ones \cite{syperek07}. This is exactly the same mechanism that in the Voigt configuration is responsible for growing peak height with growing magnetic field \cite{Kugler11}.

Having described the RSA signal features shared by both regimes, let us now point out the differences between them. First of all, similarly to the Voigt configuration, for larger decoherence rates the RSA peaks become broader (their halfwidths increase) and their heights get smaller. However, the main difference is the emergence of the inverted peaks in the weak decoherence regime for tilted magnetic fields [see Fig. \ref{fig:RSA_model}(b)]. Due to the aforementioned complex interplay of numerator and denominator of Eq. (\ref{eq:RSA_tilt}), the sign of the signal oscillations can change for magnetic fields up to some field $B_0$, which grows with the growing tilt angle.
\begin{figure}[t!]
	\begin{center}
		\includegraphics[width=\columnwidth]{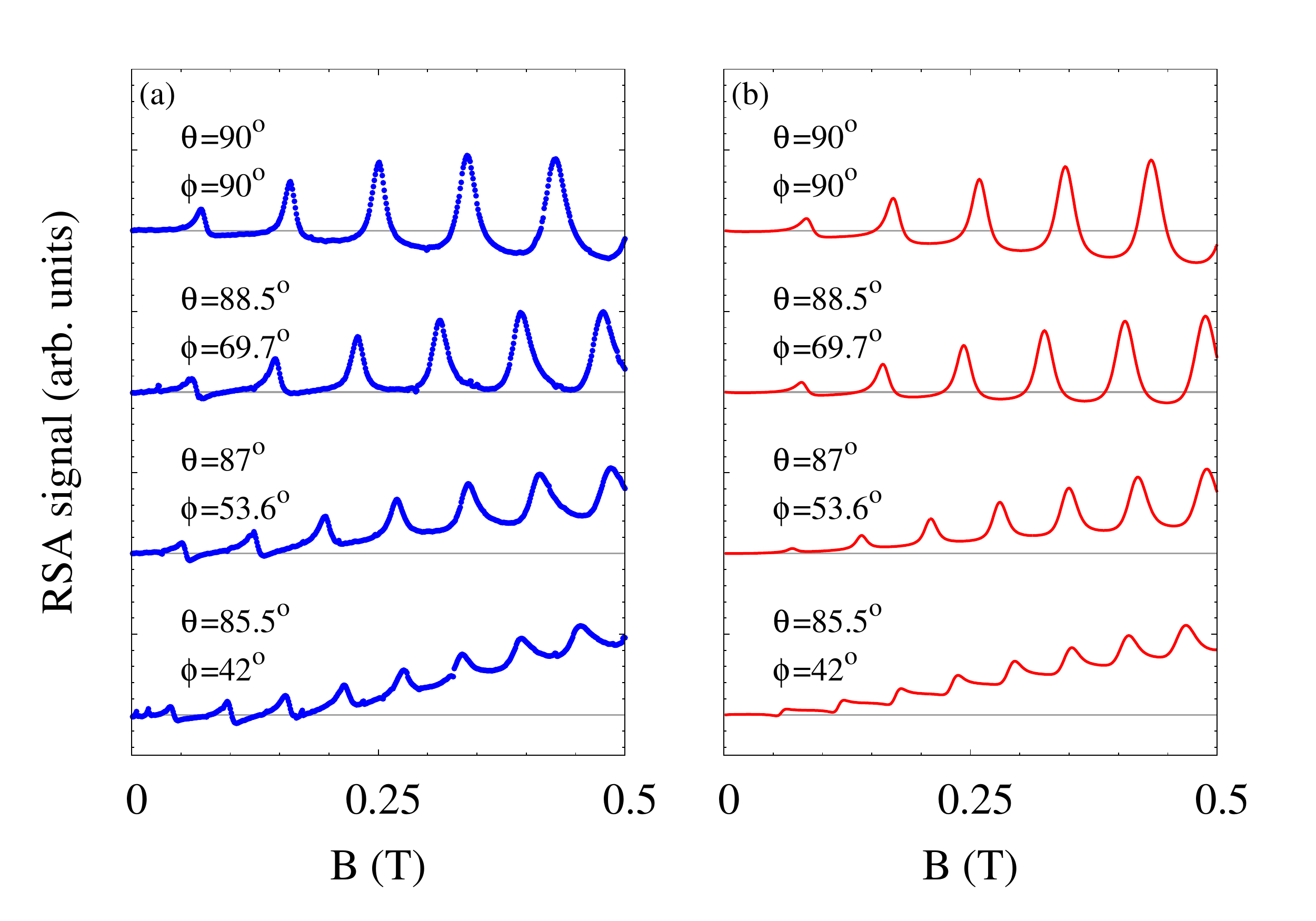}
		\caption{\label{fig:RSA_exp}(a) Experimental RSA signals in tilted magnetic fields; (b) Modeled RSA signals for tilted magnetic fields with parameters: $g_{\mathrm{t}}=0.266$, $g_{\perp}=0.066$, $\sigma_{\perp}=1.5\%\cdot g_{\perp}$, $g_{\parallel}=0.93$, $\sigma_{\parallel}=1.5\%\cdot g_{\parallel}$, $\gamma=10^{10}$ s$^{-1}$, $\kappa_x=\kappa_{x0}=\kappa_z=\kappa_{z0}=2\cdot 10^7$ s$^{-1}$.}
	\end{center}
\end{figure}

To verify that our model can reproduce the experimentally observed RSA signals in tilted magnetic fields, we performed a series of measurements. The measurements were performed for magnetic field ranging up to $0.5$ T, tilt angle varying from $0^{\mathrm{o}}$ to $4.5^{\mathrm{o}}$ and the temperature set to $T=1.2$ K. The experimental data are shown in Fig. \ref{fig:RSA_exp}(a) and the modeled signals are presented in Fig. \ref{fig:RSA_exp}(b). All the parameters of the model, apart from the tilt angle, are the same for all modeled curves. Also, to account for the inhomogeneous ensemble broadening of the hole g-factors, the modeled result was averaged according to a Gaussian distribution of g-factors, with the standard deviations $\sigma_{\perp}$ and $\sigma_{\parallel}$ for $g_{\perp}$ and $g_{\parallel}$, respectively. Although the behavior of the experimentally measured RSA signal can be reproduced using the formula given in Eq. (\ref{eq:RSA_tilt}), similarly to the TRFR case, there are too many free fitting parameters to reliably infer the values of spin dynamics parameters from the fitting procedure. Finally, we have to note that in none of our experiments have we observed the inversion of the RSA peaks, which is probably due to the hole spin decoherence rates being too large.

\section{Conclusion}
\label{sec:concl}

We have developed a theoretical description of  time-resolved Faraday rotation and resonant spin amplification experiments performed on p-doped QWs in magnetic fields tilted from the Voigt geometry. We have found analytical formulas describing both signals and identified physical processes responsible for their origin. We also predict that for long enough hole spin dephasing times it should be possible to observe inversion of the RSA peaks induced by tilted magnetic field. Our theoretical findings have been partly verified by a series of experimental measurements on a p-doped GaAs/Al$_{0.3}$Ga$_{0.7}$As single QW we have performed. Specifically, we have shown that our model can account for all the signal features observed experimentally and thus reproduce experimentally measured traces, but not all features predicted theoretically, i.e. the inversion of the RSA peaks, were observed for the sample used in the experiment.
\\

\section*{Acknowledgements}
Financial support by the DFG via SPP 1285 and SFB 689, as well as by the Foundation for Polish Science under the TEAM program, cofinanced by the
European Regional Development Fund, is gratefully acknowledged.

\bibliographystyle{prsty}
\bibliography{Nonresonant}

\end{document}